\documentclass[aps,prd,preprint,superscriptaddress,showpacs,showkeys]{revtex4}
\usepackage{amssymb,amsmath}
\usepackage{graphicx} 
\usepackage{dcolumn} 

\usepackage{epsfig}   
\usepackage{pstricks}
\usepackage{ifpdf}

\begin{document}

\title{ Effective Locality in the pure gluon sector}

\author{H.M. Fried}
\affiliation{Physics Department, Brown University, Providence, RI 02912, USA}
\email[]{herbert_fried@brown.edu}
\author{Y. Gabellini}
\affiliation{Universit\'{e} C\^ote d'Azur\\ Institut de Physique de Nice, UMR CNRS 7010 1361 routes des Lucioles, 06560 Valbonne, France}
\email[]{Yves.Gabellini@inphyni.cnrs.fr}
\author{T. Grandou}
\affiliation{Universit\'{e} C\^ote d'Azur\\ Institut de Physique de Nic{e}, UMR CNRS 7010 1361 routes des Lucioles, 06560 Valbonne, France}
\email[]{Thierry.Grandou@inphyni.cnrs.fr}

\date{\today}
\vspace{5cm}

\begin{abstract} 
About twelve years ago the use of standard functional manipulations was demonstrated to imply an unexpected property satisfied by the fermionic Green's functions of $QCD$. This non-perturbative phenomenon is dubbed \textit{Effective Locality}. In a much simpler way than in $QCD$, 
the most remarkable and intriguing aspects of Effective Locality are presented here, in the Yang--Mills theory on Minkowski spacetime. 
\end{abstract}


\maketitle


\section{Introduction}
Over the past twelve years a new property of $QCD$ has been put forth under the name of \emph{effective locality}, $EL$ for short, and can be phrased as follows,
\par

\textit{For any fermionic $2n$--point Green's functions and related amplitudes, the
full sum of cubic and quartic gluonic interactions,
fermionic loops included, results in a local contact type interaction. This local interaction is
mediated by a tensorial field which is antisymmetric both in Lorentz and
color indices. Moreover, the sum just evoked appears to be gauge invariant.}
\par
At both theoretical and phenomenological points of view, $EL$ first predictions seem interesting and mostly in line with the expected features of the $QCD$ non--perturbative regime, as well as some new predictions, more specific to the $EL$ property itself \cite{ygtg}.
\par
Now, the full $QCD$ case makes the overall $EL$ treatment quite involved ${}^{\footnotemark[1]}$,
\footnotetext[1]{In particular, a mass scale shows up necessarily in order to give a meaning to the fermionic $2n$--point Green's functions.}while the crucial issue of non--abelian gauge invariance, which would posit $EL$ as a sound property, can be read off the simpler Yang--Mills ($Y$-$M$) case. 

\par 
 This is why the current letter aims at examining the $EL$ property in the very context of the Minkowskian Yang--Mills theory, focusing on the gluonic Green's functions generating functional.  The formal context of the whole matter is that of standard functional methods within Lagrangian quantum field theories \cite{Eik}. Of course, Yang--Mills theory has been the matter of an impressive amount of work that the present letter does not pretend to list.

 \section{Gluonic Green's functions generating functional}
 \subsection{Presentation}
 In Minkowski spacetime, the $Y$-$M$ Lagrangian density reads,
 \begin{equation}\label{Lagrangien}
 {\mathcal{L}}_{Y\!M}(x)=-{1\over 4} F^a_{\mu\nu}(x)F^{a{\mu\nu}}(x),
 \end{equation}
 where the $F^a_{\mu\nu}(x)$ are the customary non--abelian field strength tensors,
  \begin{equation}
 F^a_{\mu\nu}(x)=\partial^x_{\mu}{A}_{\nu}^a(x) - \partial^x_{\nu}{A}_{\mu}^a(x) + gf^{abc}{A}_{\mu}^b(x){A}_{\nu}^c(x),
 \end{equation}
 and where the $f^{abc}$ are the structure constants of the $SU_c(3)$--Lie algebra. Out of $ {\mathcal{L}}_{Y\!M}$, the ``free'' ($g=0$) and interacting parts ($g\neq0$) can be separated, 
  \begin{equation}\label{separated}
 {\mathcal{L}}_{Y\!M}=-{1\over 4}\left(\partial_{\mu}{A}_{\nu}^a - \partial_{\nu}{A}_{\mu}^a\right)\left(\partial^{\mu}{A}^{a\nu} - \partial^{\nu}{A}^{a\mu}\right)+{\mathcal{L}}_{int.}
 \end{equation}with,
  \begin{equation}\label{int}
\mathcal{L}_{int.}[A] =  -\frac{1}{4} \left(2\, (\partial_{\mu}{A}_{\nu}^a - \partial_{\nu}{A}_{\mu}^a) +  g f^{ade} A_{\mu}^{d} A_{\nu}^{e} \right) \, \left(g f^{abc} A^{\mu}_{b} A^{\nu}_{c} \right)\end{equation}
An integration by parts yields,
 \begin{equation}\label{identic}
\int  {\mathcal{L}}_{Y\!M}=-\frac{1}{4} \int{{F}^{2}} =  \frac{1}{2} \int{ \displaystyle{A}^a_{\mu}(x)\bigl(g^{\mu\nu}\,\partial^2 - \,\partial^{\mu}\partial^{\nu}\bigr){A}^a_{\nu}(x) } + \int{{\mathcal{L}}_{int.}}[A]
\end{equation}
In a standard manner, this separation allows one to obtain the full gluonic generating functional out of the free field one :
 \begin{equation}\label{ZYM}
 Z_{YM}[j]=N\, e\,^{\displaystyle{ i\int\!\!{\mathrm{d}}^4x\,{\cal L}_{\rm int}\bigl(A^{c}_{\rho}\rightarrow{i}{\delta\over\delta j_{\rho}^c(x)}\bigr)}}\, e\,^{\displaystyle -{i\over2}\int\!\!\mathrm{d}^4x\,\mathrm{d}^4y\,\,j_{\mu}^a(x)\,D^{\mu\nu}_{ab}(x-y)\,j_{\nu}^b(y)}
 \end{equation}
 provided that the \emph{distribution} $D^{\mu\nu}_{ab}(x-y)$ exists. Now, this is the point, because $D^{\mu\nu}_{ab}$ is supposed to satisfy the equation,
 \begin{equation}\label{inverse}
 \displaystyle \bigl[g^{\mu\rho}\partial^2_x - \partial_x^{\mu}\partial_x^{\rho}\bigr]\,D_{\rho}^{ab\, \nu}(x-y) = g^{\mu\nu}\delta^{ab}\,\delta^{(4)}(x-y)
 \end{equation}
As well known, the operator $\partial^2 -\partial\otimes\partial$ cannot be inverted and $D^{\mu\nu}_{ab}$ is accordingly not defined.
 \par
 In $QED$ as well as in $QCD$, solutions to this problem have been known for long, preserving Lorentz invariance and breaking momentarily the local gauge invariance of $ {\mathcal{L}}_{Y\!M}$ by means of a gauge--fixing term.
In a second step, local gauge invariance is restored and checked through the Ward--Takahashi and Slavnov--Taylor identities satisfied by Green's functions, in the respective cases of $QED$ and $QCD$. All this is textbook material.

\subsection{Peculiarity of the non--abelian situation}
The non--abelian structure of $ {\mathcal{L}}_{Y\!M}$ offers a possibility which has no abelian equivalent. What can be done, is to add and subtract a term to (\ref{Lagrangien}), for example,
 \begin{equation}\label{plusmoins}
 {\mathcal{L}}_{Y\!M}\ \longrightarrow\   {\mathcal{L}}_{Y\!M} + \frac{1}{2\zeta}(\partial\cdot \!{A^a})^2- \frac{1}{2\zeta}(\partial\cdot \!{A^a})^2
 \end{equation}
 By doing so, nothing is changed in ${\mathcal{L}}_{Y\!M}$ and the original full gauge invariance of (\ref{Lagrangien}) is preserved, while using one of the two extra terms of (\ref{plusmoins}), the undefined expression of $D_{\mu\nu}^{ab}$ in (\ref{inverse}) is turned into the well defined covariant expression ${{D}_{\mathrm{F}}^{(\zeta)}}$, 
 \begin{equation}\label{covprop}
D_{\mu\nu}^{ab}\ \longrightarrow\  \left({{D}_{\mathrm{F}}^{(\zeta)}}^{-1}\right)_{\mu \nu}^{a b} = \delta^{a b} \, \left[ g_{\mu \nu} \, \partial^{2} - \left(1 -\frac{1}{\zeta} \right) \partial_{\mu} \partial_{\nu} \right]
\end{equation}
corresponding to the familiar expression ${}^{\footnotemark[2]}$
\footnotetext[2]{Equation (\ref{inverse}) is of course left unchanged by (\ref{plusmoins}). This is why the word \emph{expression} is used for (\ref{covprop}) and (\ref{covprop2}), rather than the denomination of `propagator'. No gauge-fixing procedure is implemented here, to which (\ref{covprop2}) would correspond as the related genuine free field propagator. See Comment (ii). }, 
 \begin{equation}\label{covprop2}\displaystyle  D_{F\mu\nu}^{(\zeta)ab}(x-y) =  -\delta^{ab}\int\!\!{d^4k\over(2\pi)^4}\,e\,^{\displaystyle -ik\!\cdot\!(x-y)}\Bigl[\,{g_{\mu\nu}\over k^2+ i\varepsilon} + \bigl(\zeta - 1\bigr)\, {k_{\mu}k_{\nu}\over(k^2+ i\varepsilon)^2}\,\Bigr] \end{equation}
\medskip
Furthermore, using (\ref{identic}) to re--express the density of interaction ${\mathcal{L}}_{int.}$, one arrives now at 
\begin{eqnarray}\label{ZYM1}
{Z}_{{YM}}[j] = {N} \, \left. e^{\displaystyle{-\frac{i}{4} \int{{F}^{2}} - \frac{i}{2}  \int{ \displaystyle{A}^a_{\mu}(x)\bigl(g^{\mu\nu}\,\partial^2 - \,(1 - \frac{1}{\zeta})\,\partial^{\mu}\partial^{\nu}\bigr){A}^a_{\nu}(x)}}} \right|_{\displaystyle A \rightarrow \frac{1}{i} \, \frac{\delta}{\delta j} } \, e^{\displaystyle{-{\frac{i}{2} \int{j \cdot {D}_{\mathrm{F}}^{(\zeta)} \cdot j} }}}\,,
\end{eqnarray}
At this stage, a convenient rearrangement of (\ref{ZYM1}) is obtained by relying on an identity satisfied by any polynomial and exponential functional $\mathcal{F}[A]$ \cite{Eik}, \begin{eqnarray}\label{reciproc}
\mathcal{F}\left[ \frac{1}{i} \frac{\delta}{\delta j} \right] \  e^{-\displaystyle{{\ \frac{i}{2} \int{j \cdot {D}_{\mathrm{F}}^{(\zeta)} \cdot j} }}} = e^{-\displaystyle{{\ \frac{i}{2} \int{j \cdot {D}_{\mathrm{F}}^{(\zeta)} \cdot j} }}} \  \left. e^{\displaystyle{{\mathfrak{D}^{(\zeta)} _{A}}}} \ \mathcal{F}[A] \right|_{\displaystyle A = \int{{D}_{\mathrm{F}}^{(\zeta)} \cdot j} }
\end{eqnarray}where the so--called ``linkage operator'' $\exp {\mathfrak{D}_{A}}$ appears,
\begin{equation}\label{linkage}
\exp\mathfrak{D}^{(\zeta)} _{A} =  \exp\frac{i}{2} \int\mathrm{d}^4x\int\mathrm{d}^4y\ {\frac{\delta}{\delta A(x)} \cdot  {D}_{\mathrm{F}}^{(\zeta)}(x-y) \cdot \frac{\delta}{\delta A(y)} }\end{equation}
In this way, one gets for (\ref{ZYM1}) another expression which reads,
\begin{eqnarray}\label{ZYM2}
{Z}_{{YM}}[j] &=& {N}\ e^{\displaystyle-{{\ \frac{i}{2} \int{j \cdot {D}_{\mathrm{F}}^{(\zeta)} \cdot j} }}}\nonumber\\ &\cdot& \left. e^{\ \displaystyle{{\mathfrak{D}^{(\zeta)} _{A}}}} \  e^{\displaystyle{{-\frac{i}{4} \int{{F}^{2}} - \frac{i}{2} \int{ A \cdot \left({ {D}_{\mathrm{F}}^{(\zeta)}}\right)^{-1} \cdot A} }}} \right|_{\displaystyle A = \int{{D}_{\mathrm{F}}^{(\zeta)}\cdot j} }
\end{eqnarray}
\par
In order to proceed, it is appropriate to introduce a ``linearization'' of the $\int F^2$ term which appears in the right hand side of (\ref{ZYM2}). This can be achieved by using the representation \cite{Halpern}, 
\begin{equation}\label{Halpern}
e^{{\displaystyle{-\frac{i}{4} \int{{F}^{2}}}}} = \mathcal{N} \, \int{\mathrm{d}[\chi] \ e^{ {\displaystyle{\ \frac{i}{4} \int{\chi_{\mu \nu}^{a}\chi^{a\mu \nu} + \frac{i}{2} \int{ \chi^{a\mu \nu} {F}_{\mu \nu}^{a}} } } }}}
\end{equation}bringing (\ref{ZYM2}) into the form,
\begin{eqnarray}\label{ZYM3}
{Z}_{{YM}}[j] &=& {N}\ e^{-\displaystyle{{\ \frac{i}{2} \int{j \cdot {D}_{\mathrm{F}}^{(\zeta)} \cdot j} }}}\int{\mathrm{d}[\chi] \ e^{ {\displaystyle{\ \frac{i}{4} \int{\chi_{\mu \nu}^{a}\chi^{a\mu \nu}}}}}}
\nonumber\\ &\cdot& \left. e^{\ \displaystyle{{\mathfrak{D}^{(\zeta)} _{A}}}} \  e^{\,\displaystyle{{\frac{i}{2} \int{\chi\cdot{F}} - \frac{i}{2} \int{ A \cdot \left({ {D}_{\mathrm{F}}^{(\zeta)}}\right)^{-1} \cdot A} }}} \right|_{\displaystyle A = \int{{D}_{\mathrm{F}}^{(\zeta)}\cdot j} }
\end{eqnarray}
The action of (\ref{linkage}) on linear and quadratic $A^a_\mu$--field dependences is established in the Appendix and yields :
\begin{eqnarray}\label{ZYM40}
{Z}_{{YM}}[j] &=& {N}\ \int{\mathrm{d}[\chi] \ e^{ {\displaystyle{\ \frac{i}{4} \int{\chi_{\mu \nu}^{a}\chi^{a\mu \nu}}}}}}\,\bigl[\det(gf\cdot\chi)\bigr]^{-\frac{1}{2}}
\nonumber\\ &\cdot&  e^{\displaystyle -{i\over2}\int\!\!\mathrm{d}^4x\,\,\left(j_{\mu}^a(x)+\partial^\lambda\chi^a_{\lambda\mu}\right)(x)\,\bigl[(gf\cdot\chi)^{-1}\bigr]^{\mu\nu}_{ab}(x)\,\left(j_{\nu}^b(x)+\partial^\sigma\chi^b_{\sigma\nu}\right)(x)}\end{eqnarray}

\subsection{Comments}
 Several comments are in order.
 \par\bigskip
 (i) Result (17) is not an {\textit{artefact}} related to the use of the functional identity (\ref{reciproc}). In order to derive ${Z}_{{YM}}[j]$, relying on (\ref{reciproc}) is in no way mandatory. We have checked that other standard functional operations may be used and yield exactly the same result. The crucial point is that these functional operations all proceed via {\em{functional differentiations}}, and not through {\em{functional integrations}} on the $A^a_\mu$--gauge field configuration space. This important difference is discussed in the next Section.
\par\medskip
(ii) Nor does (\ref{ZYM40}) depend on the tricks of (\ref{plusmoins}) (or (\ref{mass}), below).  It is important to recall that effective locality was first discovered in $QCD$ by observing that in contrast to $QED$, $QCD$ proper quantisation can be achieved by maintaining \emph{both} manifest Lorentz covariance and manifest gauge invariance :  Along this approach \cite{EPJC}, a \emph{genuine} free field Feynman propagator corresponding to $\zeta=1$ in (\ref{covprop}) and (\ref{covprop2}) is singled out; but it disappears eventually, exactly like (\ref{covprop}) and (\ref{massprop}) disappear from ${Z}_{{YM}}[j]$ in the current calculations.
\par\medskip
Adding and subtracting the terms appearing in the right hand side of (\ref{plusmoins}) is for the only purpose of generating some well defined {\em{intermediate}} expression ${D^{(\zeta)}_F}^{ab}_{\mu\nu}$, or `intermediate propagator', out of the undefined expression $D^{\mu\nu}_{ab}$ of (\ref{inverse}), while preserving gauge invariance by construction, and offering a derivation of (\ref{ZYM40}) simpler than that of Ref.\cite{EPJC}.
\par
Using (\ref{plusmoins}) or similar Lagrangians could suggest however that gauge--fixing conditions are some lurking parts of these derivations. But it is not so. This point can be illustrated very simply by endowing the gauge fields with a mass term, that is, by writing instead of  (\ref{plusmoins}),
\begin{equation}\label{mass}
 {\mathcal{L}}_{Y\!M}\ \longrightarrow\   {\mathcal{L}}_{Y\!M} +\frac{m^2}{2}\,{A_\mu^a}A^{a\mu}-\frac{m^2}{2}\,A_\mu^aA^{a\mu}.
 \end{equation}While the overall gauge invariance is preserved trivially by (\ref{mass}), as it is by (\ref{plusmoins}), the previously undefined expression $D^{\mu\nu}_{ab}$ of (\ref{inverse}) is now turned into the {\emph{intermediate}} massive Feynman propagator, ${D_F^{(m)}}^{\mu\nu}_{ab}$, with, in momentum space,
 \begin{equation}\label{massprop}
{D_F^{(m)}}^{\mu\nu}_{ab}(k)=\delta_{ab}\, \frac{-g^{\mu\nu}+k^\mu k^\nu/m^2}{k^2-m^2+i\varepsilon}.
\end{equation}Then, following the same steps as before, from (\ref{ZYM1}) to (16), the same result is obtained once again for ${Z}_{{YM}}[j]$, and this is free of any gauge-fixing considerations.
\par\noindent
Any reference to the expressions ${{D}_{\mathrm{F}}^{(\zeta,m,\dots)}}$, or `intermediate propagators', has disappeared from (17). The mechanism at the origin of this remarkable ${{D}_{\mathrm{F}}^{(\zeta,m,\dots)}}$--independence is established in details in the appendix, as a mere consequence of functional relations (A5), (A6) and (A7), operating on the specific non-abelian Yang-Mills structure: This cannot happen in the simpler abelian case.

\par\medskip
(iii) In full rigour, while the left hand side of (\ref{Halpern}) is manifestly gauge invariant, the right hand side is not. This means that a full integration of the $\chi^a_{\mu\nu}$-dependences is necessary to stick to the \emph{ab initio} guaranteed gauge invariance of the left hand side. Fortunately, in some cases at least, the powerful \emph{random matrix theory} renders this integration process a doable step \cite{ygtg}.

\par\medskip
(iv) As stated in point (i), quantisation is achieved through functional differentiation as shown in Eqs.(\ref{reciproc}), (\ref{ZYM2}) and (16), rather than through the customary functional integration where gauge--fixing conditions are used, which explicitly break gauge-invariance. In this latter case it is well known that the yet unsolved {\em{Gribov copy}} problem prevents one from reaching a proven recovering of non-abelian gauge invariance. On the contrary, it is worth emphasizing that, thanks to functional differentiation, the present construction of ${Z}_{{YM}}[j]$ is able to circumvent the Gribov copy problem. 
 
\par\medskip
(v) The striking feature of ${Z}_{{YM}}[j]$ is of course that gauge free field expressions such as ${D_F}^{ab}_{\mu\nu}(x-y)$ or ${D_F^{(m)}}^{\mu\nu}_{ab}(x-y)$ no longer show up and disappear to the exclusive benefit of the local $[(gf\cdot\chi)^{-1}]^{\mu\nu}_{ab}(x)$ expression~${}^{\footnotemark[3]}$. The unexpected locality of the effective interaction in ${Z}_{{YM}}[j]$ is one of the the non--trivial aspects of (17) on the basis of which a dual formulation of the original $Y$-$M$ theory could be sought, and proved as holding at the first non--trivial order of a semi--classical expansion \cite{RefF} (features which, aside from \emph{supersymmetric} extensions, do not generalise to the full physical $QCD$ case \cite{SW1994}).
\par\medskip

\footnotetext[3]{In Minkowski spacetime, the latter do not enjoy the interpretation they receive in the euclidean formulation, where, at leading order of a semi--classical expansion, they correspond to field strengths evaluated on {\em{instanton--field}} configurations, that is, $\chi^a_{\mu\nu}\simeq F^a_{\mu\nu}(A^{inst.})$ \cite{RefF}.}

(vi) The analytic form of ${Z}_{{YM}}[j]$ has been observed partly in \cite{RefF}, in an euclidean formulation and within a gauge-fixing procedure translated from the original $A_\mu$-gauge fields, to the new $\chi^a_{\mu\nu}$-ones. 
\par
As will be quoted elsewhere, the $j_{\mu}^a(x)$-source free ${Z}_{{YM}}[j]$ above, can be written in a form which is strongly reminiscent of a \emph{non-linear sigma model} (while not complying exactly with the most general definition of these models at $D\geq 2$, \cite{ygtg2}) .. can this be so, not by chance solely \cite{Douglas}?

\section{Conclusions}

\par
In quantum field theories, whenever the {\em{Wick theorem}} holds for chronological products of field operators, Green's functions can be calculated using either functional integrations or functional differentiations \cite{Kleinert}.  Equivalently.  The question has been addressed to know wether this equivalence holds in the non--abelian situation also \cite{tg19}. A cogent answer to this questioning is rendered uneasy because of the Gribov problem. One could think that if the latter was fully mastered, then, functional integrations would deliver the same result as functional differentiations do, that is (17) in the current case. 
\par
But the answer to the previous questioning could come out different. By ``complexifying'' the non--abelian gauge fields to be summed upon in a functional integration process, one may find several different {\em{saddle points}} of the Lagrangian action; and {\em{resurgence theory}} tells us that expansions, and resummations thereof, are connected to each others. What then would be the relation of (17) to the various expansions devised around each action saddles? These considerations are of course speculative and we have no hints about a possible answer at this stage.
\par
As noticed already in the case of $QCD$ \cite{tg19}, what appears throughout this $EL$ analysis of the $Y$-$M$ theory is a form of disconnection from {\em{Perturbation Theory}}.  The latter, in effect, requires well defined free gauge--field propagators out of which Feynman rules are devised, while nothing like this can be identified in the current treatment of the $Y$-$M$ theory. This point could be highly suggestive of a real disconnection between Perturbation Theory and the $EL$ non perturbative approaches of $Y$-$M$ and $QCD$. In this respect, it is important to recall that the Dyson--Schwinger {\em{non--truncated}} analyses of Ref.\cite{Kreimer} have been able to point out such an irreducible disconnection regarding the $\beta$--function of $QCD$; and that similar aspects of ``uneasy connection'' can be perceived in the {\em{Light-front approach}} to $QCD$ \cite{Stan}. 
\par
Likewise, it is worth noting that reaching the non--perturbative regime of $QCD$ (or $Y$-$M$) theories from a {\em{BRST--quantisation procedure}} is suspected to be hardly doable \cite{becchi}, and to let the analytic properties of $QCD$ propagators ``largely unknown'' \cite{Lowdon}. 
\par
Roughly speaking, one could venture the idea that $BRST$--quantised $QCD$ could trap $QCD$ within its perturbative regime.
\par
It remains to take advantage of (17) in the course of concrete calculations. At a first glance, it seems that in order to make sense, the Green's functions generated out of (\ref{ZYM40}) call necessarily for a mass, like in the full $QCD$ case \cite{ygtg2}.

\bigskip

\appendix

{\section{On the $D_F$--independence of $Z_{YM}[j]$ }
\setcounter{equation}{0}

One starts from (16--\ref{ZYM3}) :

\begin{eqnarray}\label{ZYM3}
{Z}_{{YM}}[j] &=& {N}\ e^{-\displaystyle{{\ \frac{i}{2} \int{j \cdot {D}_{\mathrm{F}}^{(\zeta)} \cdot j} }}}\int{\mathrm{d}[\chi] \ e^{ {\displaystyle{\ \frac{i}{4} \int{\chi_{\mu \nu}^{a}\chi^{a\mu \nu}}}}}}
\nonumber\\ &\cdot& \left. e^{\ \displaystyle{{\mathfrak{D}^{(\zeta)} _{A}}}} \  e^{\,\displaystyle{{\frac{i}{2} \int{\chi\cdot{F}} - \frac{i}{2} \int{ A \cdot \left({ {D}_{\mathrm{F}}^{(\zeta)}}\right)^{-1} \cdot A} }}} \right|_{\displaystyle A = \int{{D}_{\mathrm{F}}^{(\zeta)}\cdot j} }
\end{eqnarray}

Now, 
 \begin{equation}\chi\cdot{F} = {\chi}_{\mu\nu}^a(x)\,{F}^{a\mu\nu}(x) = -2\, (\partial^{\nu}{\chi}_{\nu\mu}^a(x))\,{A}_a^{\mu}(x) + {A}_a^{\mu}(x)\,gf^{abc}{\chi}_{\mu\nu}^c(x){A}_b^{\nu}(x)
\end{equation}
so that :
        \begin{eqnarray}{\label{ZYM4}}\displaystyle  {Z}_{{YM}}[j] & =&N\,\displaystyle e\,^{\displaystyle -{i\over2}\int\!\!{\mathrm{d}}^4x\,{\mathrm{d}}^4y\,j^{\mu}_a(x)\,D_{F\mu\nu}^{(\zeta)ab}(x-y)\,j^{\nu}_b(y)}\,\int\!\!d[\chi]\,e\,^{\displaystyle  {i\over 4}\int\!\!{\mathrm{d}}^4x\, ({\chi}_{\mu\nu}^a(x))^2}\nonumber\\ &{}& \displaystyle \,e\,^{\displaystyle {i\over2}\int\!\!{\mathrm{d}}^4x\,{\mathrm{d}}^4y\,{\delta\over\delta A^a_{\mu}(x)}\,D_{F\mu\nu}^{(\zeta)ab}(x-y)\,{\delta\over\delta A^b_{\nu}(y)}}\,e\,^{\displaystyle{-i}\int\!\!{\mathrm{d}}^4x\, (\partial^{\nu}{\chi}_{\nu\mu}^a(x) ){A^a}^{\mu}(x)}\\ &\times&\,\,e\,^{\displaystyle{-{i\over 2}\int\!\!{\mathrm{d}}^4x\,{A}_a^{\mu}(x)\Bigl[\bigl(g_{\mu\nu}\,\partial^2 - (1 - {1\over\zeta} )\,\partial_{\mu}\partial_{\nu}\bigr)\delta^{ab} - ( g\,f\!\cdot\!\chi)^{ab}_{\mu\nu}(x)\Bigr]{A}_b^{\nu}(x)}}\Biggl|_{\displaystyle A =\int \!\!D\cdot j}\nonumber\hfill
\end{eqnarray}
The action of the linkage operator upon a gaussian functional yields \cite{Eik} :
\begin{eqnarray}
&{}&\displaystyle e\,^{\displaystyle {i\over2}\!\int\!\!{\delta\over\delta A}D{\delta\over\delta A}}\,\,e\,^{\displaystyle -{i\over2}\!\int\!\!ABA -{i}\!\int\!\!FA}\\ &{}&= \displaystyle e\,^{\displaystyle -{i\over2}\!\int\!\!AB(1 - DB)^{-1}A}\,e\,^{\displaystyle -{i}\!\int\!\!F(1 - DB)^{-1}A}\,e\,^{\displaystyle -{i\over2}\!\int\!\!F(1 - DB)^{-1}DF}e\,^{\displaystyle -{1\over2}{\rm Tr}\ln(1-DB)}\nonumber
\end{eqnarray}
With
 ${\displaystyle A =\int \!\!D\cdot j,\  B = D^{-1} - gf\cdot\!\chi,\  F = \partial\cdot\!\chi}$, one gets :
\begin{equation}AB(1 - DB)^{-1}A = j\cdot\!(gf\cdot\!\chi)^{-1}\cdot\!j - j\cdot\!D\cdot\!j\end{equation}
\begin{equation}F(1 - DB)^{-1}A = \partial\cdot\!\chi\,(gf\cdot\!\chi)^{-1}\cdot\!j\end{equation}
\begin{equation}F(1 - DB)^{-1}DF = \partial\cdot\!\chi\,(gf\cdot\!\chi)^{-1}\,\partial\cdot\!\chi\end{equation}
which leads right away to (17). As mentioned in the main text, other standard functional methods lead to the very same result, so long as they proceed along functional differentiation.

\end{document}